\documentclass{iopart}
\usepackage{graphicx}
\usepackage{color}
\begin{document}
\title[Search algorithm for GWs from GRB030329]{Search algorithm for a gravitational wave signal
in association with Gamma Ray Burst GRB030329 using the LIGO detectors}
\author{S.~D.~Mohanty$^\dag$, Sz.~M\'arka$^\ddag$, R.~Rahkola$^\S$ , S.~Mukherjee$^\dag$, I.~Leonor$^\S$ , R.~Frey$^\S$,  J.~Cannizzo$^\P$, J.~Camp$^\P$ }
\address{$^\dag$ The University of Texas at Brownsville and Texas Southmost College, Brownsville, TX 78520, USA.}
\address{$^\ddag$ LIGO -- California Institute of Technology, Pasadena, CA 91125, USA.}
\address{$^\S$ University of Oregon, Eugene, OR 97403, USA.}
\address{$^\P$ NASA/Goddard Space Flight Center, Greenbelt, MD 20771, USA.}
\begin{abstract}
One of the brightest Gamma Ray Burst ever recorded, GRB030329,
occurred during the second science run of the LIGO detectors. At
that time, both interferometers at the Hanford, WA LIGO site were
in lock and acquiring data. The data collected from the two
Hanford detectors was analyzed for the presence of a gravitational
wave signal associated with this GRB. This paper presents a
detailed description of the search algorithm implemented in the
current analysis.
\end{abstract}
\pacs{}
\submitto{\CQG}
\section{Introduction}
On 29 March 2003, one of the closest and brightest ever Gamma Ray
Burst, GRB030329~\cite{Vanderspek+etal:2003}, was detected by
instruments aboard the HETE-2 satellite.  Subsequent afterglow
observations fixed the redshift of the host galaxy at $z\simeq
0.17$~\cite{Greiner+etal:2003}. At the time when the GRB030329 was
detected, the two LIGO interferometers at the Hanford site, were
in lock and taking data. The data in the vicinity of the GRB
trigger time was analyzed for the presence of a gravitational wave
burst signal. The results from the analysis as well as relevant
details about the trigger itself will be reported
elsewhere~\cite{GRB030329_LSC}. In this paper we present details
of the search algorithm used in this analysis.

The distinction between a full analysis pipeline and a search
algorithm should be emphasized. The analysis pipeline is the
entire process that starts with the raw interferometric data
output and ends with a statistical inference, such as a detection
or no-detection, or a confidence interval estimate.

The search algorithm is the component which acts on pre-processed
data and produces a list of candidate events
 or the value of a statistic
which is then used in drawing a statistical inference. A first
version of a full analysis pipeline was reported
in~\cite{Amaldi03} and the pipeline used for the present analysis
will be described in~\cite{GRB030329_LSC}.

The paper is structured as follows. Section~\ref{overview}
presents the line of reasoning that leads to the current form of the
search algorithm.
Section~\ref{preprocessing} gives a brief outline of the
preparation of raw data before it is fed to the algorithm.
Section~\ref{calculation} then describes the algorithm at a level
of detail which should be sufficient to allow independent
reproduction. Our conclusions are presented in
Section~\ref{conclusions}.


\section{Justification of the algorithm}
\label{overview}

The majority of present day astrophysical models
and theories of GRBs suggest that the associated gravitational
wave signal is a short burst lasting for tens of milliseconds
contemporaneous with the GRB. There are also some models which
predict signals lasting over tens of seconds. The analysis carried
out in~\cite{GRB030329_LSC} focusses on short duration
 ($< 150$~msec) bursts.

At the time GRB030329 occurred,  only the two co-located and
co-aligned Hanford detectors were acquiring data. The spectral
responses of these detectors are also very similar. Therefore, the
arrival time and measured waveforms of a gravitational wave signal
should be essentially the same in both detectors. Ergo, it is
natural to consider cross-correlation of the two data streams as
the basis of a search algorithm. This conclusion can also be
reached via a more formal argument based on the maximum
log-Likelihood ratio test.

Let us use a simple, albeit unrealistic, model for detector noise
to illustrate the rationale behind the method. Assume that the two
detectors have uncorrelated noise. Further assume that the noise
in each is zero mean, Gaussian and stationary. Without loss of generality,
this is equivalent to the noise being white with unit
variance. Let the GW signal waveform in the output of H1 be ${\bf
h}=\{h[0],h[1],\ldots,h[N-1]\}$,  where $N$ is the length of the
data segment being analyzed. The signal waveform in the output of
H2 will be essentially the same modulo an overall amplitude factor
that we will ignore for the present argument.

Let ${\bf x}_1=\{x_1[0],x_1[1],\ldots,x_1[N-1]\}$ and
${\bf x}_2=\{x_2[0],x_2[1],\ldots,x_2[N-1]\}$ denote
data segments from H1 and H2 respectively.
For the case outlined above, the Logarithm of the Likelihood Ratio~\cite{Helstrom}
 is given by,
\begin{eqnarray}
\Lambda({\bf x}_1, {\bf x}_2 | {\bf h}) &=& {\rm ln}\left[
\frac{p({\bf x}_1,{\bf x_2}|{\bf h})}{p({\bf x}_1,{\bf x_2}|{\bf 0})}
\right] \;,\nonumber\\
&=& \sum_{i=0}^{N-1} x_1[i]h[i] +
\sum_{i=0}^{N-1}x_2[i]h[i]
-\sum_{i=0}^{N-1} h[i]^2\;,
\end{eqnarray}
where $p({\bf x}_1,{\bf x_2}|{\bf h})$ is the joint probability density of the
data in the presence of a signal $\bf h$.

$\Lambda({\bf x}_1, {\bf x}_2|{\bf h})$ is parametrized by the
signal sample values $h[k]$. In the absence of any prior knowledge
about the signal waveform, we shall simply maximize $\Lambda({\bf
x}_1, {\bf x}_2|{\bf h})$ over each $h[k]$ independently. The
maximum value, $\lambda({\bf x}_1, {\bf x}_2)$, of $\Lambda({\bf
x}_1, {\bf x}_2|{\bf h})$ occurs for ${\bf h}=({\bf x}_1+{\bf
x}_2)/2$, and
\begin{equation}
\lambda({\bf x}_1, {\bf x}_2) = \frac{1}{4}\sum_{i=0}^{N-1} x_1[i]^2
+\frac{1}{4}\sum_{i=0}^{N-1}x_2[i]^2
+\frac{1}{2}\sum_{i=0}^{N-1}x_1[i] x_2[i]\;.
\end{equation}
Thus, the test statistic obtained from the Likelihood Ratio
prescription involves the second moments
 of the individual data segments
  and their cross-correlation.

Our simplified assumptions do not hold for real data:
interferometer noise is non-Gaussian and non-stationary; H1 and H2
are not completely uncorrelated. Among sources of deviation from
Gaussianity, stationarity and independence are high power
narrowband noise (also called {\em lines}) and \emph{bursts} of
instrumental origins.

Lines in the two data streams with nearby frequencies can produce
large spurious correlations, especially in short data segments.
The presence of bursts will adversely affect the performance of
$\lambda({\bf x}_1, {\bf x}_2)$ since even if such a non-GW burst
were to occur in only one detector, $\lambda({\bf x}_1, {\bf
x}_2)$ may cross the detection threshold because of the variance
terms.

Lines are probably the easier problem to address, 
since most high power features seems to be persistent in their
carrier frequencies and bandwidths. Therefore, they are easily
identified and can be removed using notch filters (see
Section~\ref{preprocessing}). Lines that are not present at all
times pose a more complex problem but this is still easier to
trace than  instrumental bursts. Therefore, it makes sense to use
only the cross-correlation term as the basis of a detection
strategy. (A cross-correlation based waveform consistency test has
also been independently developed for use with untriggered burst
searches~\cite{cadonati:2003}.)

We know that GW signals must have about the same time of arrival
in the output of  the two Hanford detectors but we do not know
{\em a priori} the delay between the GRB and the GW signal.
Consequently, we do not know the absolute time of arrival of the
GW signal itself. However, based on most present astrophysical
models and known observational uncertainity in the onset of a GRB,
we can make an educated guess about the start and stop times of
the data segment to be analyzed. In our analysis, the start time
is chosen to be 120~sec before the GRB arrival time and the stop
time is fixed at 60~sec after it. We also verify the insensitivity
of the search for 20-30\% perturbation of these boundaries. This
interval ($\sim$180s), called the {\em on-source} segment, is much
larger than the expected duration of GW signal
($\sim$O(10-100ms)).

The optimum length of the data segments to use for computing the
cross-correlation depends on the duration of the  signal and its
signal-to-noise ratio, both of which are not known {\em a priori}.
The combination of these two facts suggests that the
cross-correlation of H1 and H2 outputs should be computed  on
segments whose start times and lengths vary over a range covering
the allowed range of arrival time and the expected duration of the
GW burst signal.

An artificial time shift, much larger than the expected signal
duration of $\sim$O(10-100ms), between the two data streams before
computing cross-correlations suppresses the average contribution
from the GW signal. This fact can be used to make a local
estimation of  noise properties, thus mitigating the effects of
non-stationarity in the interferometer outputs.

The line of reasoning outlined above naturally leads to a search
algorithm~\cite{Amaldi03} that processes the data as follows. (1)
A three dimensional random field is constructed, consisting of
cross-correlations for a range of start times, integration lengths
and time shifts. A signal must produce a local change in the
distribution of samples in this random field. (2) The signature of
the signal is extracted in a systematic manner. First, the three
dimensional field is reduced to a two dimensional image, called a
{\em corrgram}. The value of a pixel in the corrgram indicates the
significance of  cross-correlations near zero time shift relative
to large time shift cross-correlations. (3) Finally, a list of
events is found by identifying significant regions in the corrgram
image, each event being characterized by a peak time, optimal
integration length and an event strength. The strength of each
event can be compared with a preset detection threshold or can be
translated into a confidence interval/upper limit for the strength
of the underlying GW signal.

The detection threshold and the confidence intervals are obtained
via extensive Monte Carlo simulations using sections of the interferometer
output, called {\em off-source} data, that are well outside the on-source segment.

\section{Pre-processing and data conditioning}

\label{preprocessing} To mitigate the effect of instrumental
artifacts and uneven spectral sensitivity, the raw output from the
detectors needs to be processed before using the search algorithm.
 A brief description of this part of the pipeline  is provided
 here for the sake of completeness.

The output from each interferometer is broken into 330~sec long
sections with a 15 second overlap between consecutive segments,
providing a continuous tiling of the data with 300s long segments.
In order to avoid edge effects, the on-source segment (180~sec
long), lies in the middle of one such 330~sec long segment. The
pre-processing steps are, as follows. For each 330~sec long
segment:

\begin{enumerate}

\item Time-domain, zero-phase notch filters are applied to remove
a set of narrow frequency bands corresponding to known line
features.

\item A simple IIR filter is applied to H2 data with a phase
 response that  approximates the difference in the phase
responses of H1 and H2 calibration transfer functions. This
restores the proper phase and waveform relation between the
possible gravitational wave signals in the two data streams. This
is to compensate for the slightly different spectral response of
the two Hanford interferometers.

\item Each section is bandpass filtered using time domain
zero-phase filtering. The lower and upper frequencies of the pass
band are 80~Hz and 2048~Hz respectively.

\item The bandpassed data is downsampled from 16384~Hz to 4096~Hz
by dropping the appropriate subset of samples.

\item A linear predictor filter~\cite{Chatterji} (LPF) is trained
on the data between 5 and 15 seconds at the beginning. The LPF is
a procedure for constructing a high order FIR filter whose
transfer function is the inverse of the square root of the Power
Spectral Density (PSD) of the data.

\item The first and last 5~sec of the resulting data are dropped
in order to avoid filter transients. Only very short portions of
first and last 10~sec are processed by the search algorithm, in
order to avoid edge effects at the beginning and the end of each
300s long tile.

\end{enumerate}

The result of the pre-processing is a 320~sec segment consisting
of noise with a flat PSD ({\em white} noise ) between
 $\sim$80~Hz and $\sim$2048~Hz.
In practice, only the middle 300 sec from each 320~sec segment are
used to find events. The first and last 10 seconds are not used to
make the final event list but are processed in order to take care
of edge effects.

The preprocessing and data conditioning outlined above also
modifies a signal present in the data. However, all the filters
applied to the data, except for the H2 spectral shape correction
and the LPF derived filter, are the same for both H1 and H2 and
zero-phase filtering is used for all filters, including the LPF
derived ones. The spectral shape of H1 and H2 were close enough
that the LPF whitening filters did not have to be significantly
different. This ensured that the signal start times in the
pre-processed data had no significant shift and the signal phases
and waveforms remained about the same in the pre-processed and
conditioned data streams.


\section{Detailed description of analysis algorithm}
\label{calculation} We will now describe the main analysis
algorithm that is applied to pairs of pre-processed
data segments from H1 and H2. The basic idea behind the algorithm has
already been stated in Section~\ref{overview}.

First, we fix our notation.
\begin{enumerate}
\item Let the pre-processed time series from H1 be ${\bf H}_1=\{H_1[0],H_1[1],\ldots H_1[N-1]\}$
and from H2 be ${\bf H}_2=\{H_2[0],H_2[1],\ldots H_2[N-1]\}$.
\item Let ${\bf s}_1[k,j]$ be a subsegment from ${\bf H}_1$, ${\bf s}_1[k,j] = \{
H_1[k],H_1[k+1],\ldots,H_1[k+j-1]
\}$. Similary define ${\bf s}_2[k,j]$ be a subsegment from ${\bf H}_2$.
\item The cross-correlation $C_{k,m,j}$ between segments ${\bf s}_1[k,j]$ and
${\bf s}_2[m,j]$ is defined as,
\begin{equation}
C_{k,m,j}= \sum_{i=0}^{j-1} H_1[k+i] H_2[m+i]\;.
\end{equation}
(Note that the ordering of the first two indices in the definition
of $C_{k,m,j}$ is important since they correspond to data from
different detectors.) \item The length $j/4096$~sec of the segment
pair is the {\em integration length} and $k/4096$~sec is the {\em
offset}. \item Let ${\cal C}_{k,j}[p]$, $p = 0,\pm1,\pm2,\ldots$,
be the sequence,
\begin{equation}
{\cal C}_{k,j}[p] = \left[C_{k-j/2,k+p-j/2,j}^2 + C_{k-p-j/2,k-j/2,j}^2\right]^{1/2}\;.
\end{equation} ($j$ is always kept even.)

${\cal C}_{k,j}[p]$ contains the autocorrelation of the
 GW signal for samples at and
near $p=0$, while far away samples only have the signal-noise
cross-correlation. The far from zero time shift samples of ${\cal
C}_{k,j}[p]$ can be used to make an estimation of background mean
and variance. This estimate uses detector noise that is localized
near $p=0$ and is, thus, advantageous in the presence of
non-stationarity.

\item We divide the sequence ${\cal C}_{k,j}[p]$ into two parts:
{\bf
\begin{equation}
\left\{
\begin{array}{ll}
-{\cal W}/2\leq p \leq {\cal W}/2  & core \\
& \\
2.5j\leq |p| \leq 2.5 j + {\cal W} & side\; lobes\\
\end{array}
\right .
\end{equation}
}
\end{enumerate}
where ${\cal W}$ is an even integer. For a core size ${\cal W}$
comparable to the effective duration of the signal autocorrelation
function, a sum over core samples will have a comparable or larger
signal to noise ratio than the sample at $p=0$. In the analysis
algorithm, the core size is $\max\{20,j\}\leq {\cal W} \leq
\min\{j,40\}$. In practice, the analysis is found not to be
sensitive to the exact choice of the core size around this point.

The steps in the algorithm are as follows. First, we fix a
rectangular grid in offset and integration length space. Let the
list of integration lengths be $I$ and that of offsets be $O$. Let
the coordinates of the grid points be $(k,j)$, with
$I[k]/4096$~sec and $O[j]/4096$~sec being the integration length
and offset respectively. The spacing, $(O[k+1]-O[k])/4096$,
between grid points along the offset direction is $1$~msec. The
distribution of the integration lengths are logarithmic. Each
consecutive integration length is 50\% longer than the previous
one.
\begin{enumerate}

\item For each  point $(k,j)$, the sequence ${\cal C}_{O[j],I[k]}[p]$ is
computed over a range of $p$ covering the lobes.

\item The mean $\widehat{\mu}_{k,j}$ and standard deviation
$\widehat{\sigma}_{k,j}$ of ${\cal C}_{O[j],I[k]}[p]$ in the lobe
regions is computed.

\item The values of ${\cal C}_{O[j],I[k]}[p]$ in the core region
are standardized by subtracting $\widehat{\mu}_{k,j}$ and then
dividing by $\widehat{\sigma}_{k,j}$. {\em Positive} standardized
values in the core region are summed. Let the sum be $X_{k,j}$. We
call this two dimensional random field the {\em corrgram} and each
pixel contains the ``excess correlation" of close-to zero versus
non-zero time shifts. (A standardization using unphysical time
shifts was also used in~\cite{modestino+moleti:01}.)

\item \label{standardize_corrgram} For each $j$,

\begin{enumerate}

\item \label{drop_outliers} the largest 3\% of values are dropped
from the sequence $X_{k,j}$.

\item A Gamma distribution fit is found for the distribution of
the uncorrelated portion
 of the remaining pixels. The mean
$\mu_\gamma[j]$ and variance $\sigma^2_\gamma[j]$ are computed and
stored.

\end{enumerate}

\item Each $X_{k,j}$ is standardized by subtracting
$\mu_\gamma[j]$ and dividing by $\sigma_\gamma[j]$. \\

The purpose of step~\ref{standardize_corrgram} is to bring all
pixels of the corrgram on to the same footing. The purpose of
step~\ref{drop_outliers} is to prevent the biasing of the
significance estimate by the presence of a strong signal or
instrumental artifacts.

\item A very low threshold of 0.05 is used to partition the
standardized corrgram into pixels with values above and below the
threshold. The pixels in the latter set are assigned the value
zero. We continue to denote each pixel value as $X_{k,j}$.

\item A clustering algorithm is used to identify groups of adjacent
pixels with non-zero values.
The steps in the clustering algorithm are as follows:
\begin{enumerate}
\item Each pixel $X_{k,j}$ is assigned a binary score,
\begin{equation}
\begin{array}{ll}
\mbox{score 1 if} & X_{k\pm 1,j\pm 1}\neq 0\;,
\end{array}
\end{equation}
and score 0 otherwise. For pixels at the borders of the corrgram, only
the five nearest neighbours are required to be non-zero.
\item For all $(k,j)$ such that $X_{k,j}$ has score 1, set $X_{k,j}=0$ if
$X_{k,j}<0.5$. We thus drop a group of pixels whose innermost
pixel has a low value. This reduces the number of small clusters due to background
noise. The likelihood of rejecting a genuine signal is small since, for the corresponding
cluster of pixels, higher pixel values are most likely to occur inside the cluster and
not on its border.
\end{enumerate}
\item Take $\Lambda=\max_{k,j} X_{k,j}$ and let $\Lambda = X_{m,n}$ ($m,n$ thus being the
location of the maximum).\label{cluster_loop1}.
\item Let $G_{m,n}=\{X_{k,j}\;|\; n-b_0 \leq j\leq n+b_0\}$, where $b_0$ is such that
$|O[n\pm b_0]-O[n]|/4096 =
0.064$~sec, that is, $G_{m,n}$ is the set of all pixel values within $64$~msec of the
maximum. (Border pixels are handled as special cases.) Thus, we count pixels lying within
64~msec of each other as a single event. This choice is guided by Monte Carlo studies using
a variety of waveforms.
\item The
average of the 5 highest pixels in $G_{m,n}$ is computed. This quantity is called
the {\em strength} of the event. The triplet of strength, integration length $=I[m]/4096$~sec
and time of arrival $=O[n]/4096$~sec are stored. Each triplet constitutes an {\em event}.
\item Set $X_{k,j}=0$ for $n-b_0\leq j \leq n+b_0$. Repeat step~\ref{cluster_loop1} until
no more non-zero pixels are left.
\item Finally a
list of {\em events} is obtained. Each event is a potential GW signal candidate.
\end{enumerate}
The list of events is used as the basis for statistical inferences
which form a separate part of the analysis pipeline. In brief, two
types of inferences are drawn, detection and, in the absence of
detection, an upper limit on the root sum square of a possible
gravitational wave signal. For the former, a detection threshold
on the strength of each event, corresponding to a reliably known
false alarm rate, is used. The threshold is estimated via Monte
Carlo simulations using off-source data. The upper limit
estimation is also performed via Monte Carlo simulations based on
off-source data but this time with a variety of signal waveforms
injected in software. The root sum square value of a signal that
leads to, say, 90\% detection probability for a given threshold is
taken as the 90\% confidence level upper limit on the root sum
square of a possible GW signal. Details are provided
in~\cite{GRB030329_LSC}.

\section{Conclusions}
\label{conclusions} We have presented a detailed description of
the algorithm used in a search for gravitational waves associated
with GRB030329 using data from the two Hanford LIGO
interferometers. The search algorithm involves a scan  over the
all three free parameters involved in the cross-correlation based
search: integration length, offset and relative time-shifts. This
is followed by a systematic extraction of the information
contained in the resulting three dimensional phase space in the
form of an event list.

The performance of the full analysis pipeline has been quantified
using simulated data and the results are encouraging. As far as
the search algorithm itself is concerned, more effort will be put
towards the analytical characterization of its performance. We
will also explore improvements in the computational performance of
the code. This is required in order to reduce the time needed for
the background studies and Monte Carlo simulations which dominates
computational costs at the present.

\ack
We thank L.~Cadonati for thoughtful comments on the text of the paper.
The authors gratefully acknowledge the support of the United
States National Science Foundation for the construction and
operation of the LIGO Laboratory and the Particle Physics and
Astronomy Research Council of the United Kingdom, the
Max-Planck-Society and the State of Niedersachsen/Germany for
support of the construction and operation of the GEO600 detector.
The authors also gratefully acknowledge the support of the
research by these agencies and by the Australian Research Council,
the Natural Sciences and Engineering Research Council of Canada,
the Council of Scientific and Industrial Research of India, the
Department of Science and Technology of India, the Spanish
Ministerio de Ciencia y Tecnologia, the John Simon Guggenheim
Foundation, the David and Lucile Packard Foundation, the Research
Corporation, and the Alfred P. Sloan Foundation.


\section*{References}
\begin{thebibliography}{99}
\bibitem{Vanderspek+etal:2003} Vanderspek R \etal 2003 {\em GCN circular} 1997
\bibitem{Greiner+etal:2003} Greiner J \etal 2003 {\em GCN circular} 2020
\bibitem{GRB030329_LSC} LIGO Science Collaboration : Abbot B \etal 2004 {\em To be submitted}.
\bibitem{Amaldi03} Mohanty S D \etal 2004 {\em Class. Quantum Grav.} {\bf 21} S765-S774
\bibitem{Helstrom} Helstrom C W 1968 {\em Statistical Theory of Signal
Detection} 2nd. ed (London: Pergamon Press)
\bibitem{Burstpaper} LIGO Scientific Collaboration: Abbott B \etal 2004, {\em Phys. Rev. }D
 Accepted for publication\\
(LIGO Scientific Collaboration: Abbott B \etal 2004 {\em Preprint} gr-qc/0312056 )
\bibitem{cadonati:2003} Cadonati L 2004 {\em Submitted to Proc. $8^{\rm th}$
 Gravitational Wave Data Analysis
Workshop (Milwaukee)} ({\em Class. Quantum Grav.})
\bibitem{Chatterji}Chatterji S 2004 {\em LIGO technical document} LIGO-G040006-00-Z
\bibitem{modestino+moleti:01} Modestino G, Moleti A 2001 \PR D {\bf 65} 022005
\end{thebibliography}
\end{document}